\documentclass[10pt]{article}
\usepackage{amsmath}
\usepackage{cite} 
\usepackage{indentfirst}
\usepackage{latexsym}
\usepackage{pstricks}

\long\def\ca#1\cb{} 

\newcommand{\AND}{{\small AND}}

\newcommand{\becs}{\begin{cases}}
\newcommand{\bem}{\begin{matrix}}

\newcommand{\coln}{\hbox{:}}

\newcommand{\dya}[1]{|#1\rangle\langle#1|}

\newcommand{\encs}{\end{cases}}
\newcommand{\enm}{\end{matrix}}

\newcommand{\ket}[1]{|#1\rangle }

\newcommand{\msk}{\medskip }

\newcommand{\od}{\odot }

\newcommand{\ot}{\otimes }

\newcommand{\ra}{\rightarrow }

\newcommand{\st}{\sqrt{2}}

\usepackage[bf,small,center]{titlesec}
\titlespacing{\section}{1pc}{*2}{*1}

\def\ML{\mbox{ML}}\def\MR{\mbox{MR}}

\begin{document}

\centerline{\Large Quantum Counterfactuals and Locality}

\msk

\centerline{Robert B. Griffiths}
\centerline{Physics Department}
\centerline{Carnegie-Mellon University}
\vspace{.2cm}

\centerline{Version of 26 December 2011}
\vspace{0.5 cm}


\ca
\title{Quantum Counterfactuals and Locality}

\author{Robert B. Griffiths
\thanks{Electronic mail: rgrif@cmu.edu}\\ 
Department of Physics,
Carnegie-Mellon University,\\
Pittsburgh, PA 15213, USA}
\date{Version of 15 August 2011}
\maketitle  
\cb

\begin{abstract}
  Stapp's counterfactual argument for quantum nonlocality based upon a Hardy
  entangled state is shown to be flawed. While he has correctly analyzed a
  particular framework using the method of consistent histories, there are
  alternative frameworks which do not support his argument. The framework
  dependence of quantum counterfactual arguments, with analogs in classical
  counterfactuals, vitiates the claim that nonlocal (superluminal) influences
  exist in the quantum world.  Instead it shows that counterfactual arguments
  are of limited use for analyzing these questions.
\end{abstract}


\section{Introduction}
\label{sct1}

Henry Stapp \cite{ntk01} has challenged an argument by the author
\cite{Grff11,ntk02} which claims to demonstrate that quantum mechanics is a
local theory, in the sense that it contains no long-range dynamical
influences.  Stapp asserts that the validity of a certain counterfactual
statement, SR in Sec.~\ref{sct4} below, referring to the properties of a
particular particle, depends upon the choice of which measurement is made on a
different particle at a spatially distant location. He further claims that
this dependence is confirmed by an approach to counterfactuals presented in
Ch.~19 of the author's \textit{Consistent Quantum Theory} \cite{Grff02c},
hereafter referred to as CQT.
It will be argued that, on the contrary, the possibility of deriving the
counterfactual SR depends on the point of view or perspective that is
adopted---specifically on the \emph{framework} as that term is employed in
CQT---when analyzing the quantum system, and this
dependence makes it impossible to construct a sound argument for
nonlocality, contrary to Stapp's claim. 

The study below is based upon the (``consistent'' or ``decoherent'')
\emph{histories} formulation of quantum mechanics.  Since there is an extended
treatment in CQT, a brief summary in \cite{Grff09b}, and expositions of
intermediate length in \cite{Hhnb10,Grff11b}, the details need not be repeated
here, though it is worth emphasizing some features of this approach.  First,
in contrast to standard textbook quantum mechanics the histories approach
permits a discussion of \emph{microscopic} quantum properties corresponding to
subspaces of the quantum Hilbert space \emph{without} any reference to
measurements.  As a result it has no measurement problem of the sort that
besets standard quantum mechanics, which in its more orthodox versions is only
allowed to speak about macroscopic preparations and measurement outcomes.
Microscopic and macroscopic systems are addressed in the histories approach
using exactly the same fundamental quantum principles, and one can explain how
and why a measurement outcome (pointer position) can be related to the
microscopic quantum property that precedes the measurement, the property that
the measurement is intended to reveal.  Because of its ability to relate
measurement outcomes to prior measured properties the histories approach is
sometimes thought to be a ``hidden variables'' theory, and then dismissed as
necessarily leading to the usual paradoxes associated with such theories.  But
it is \emph{classical} hidden variables which, as Stapp points out, are
inconsistent with quantum principles and lead to paradoxes, whereas the
histories approach employs genuine \emph{nonclassical} microscopic (and
macroscopic) quantum properties, which are treated in a consistent way that
avoids paradoxes.

Counterfactuals and their use in quantum theory are discussed in Secs.~19.3
and 19.4 of CQT.  A brief introduction, which should suffice for the present
paper, is given in Sec.~\ref{sct2} below, based upon a particular (classical)
example.  Stapp's criticism in \cite{ntk01} uses a quantum state introduced by
Hardy \cite{Hrdy93}. That state and some associated measurements are presented
in Sec.~\ref{sct3} using the notation for entangled qubits (spin-half or
two-state systems) common in current discussions of quantum information.
Stapp's counterfactual SR is the topic of Sec.~\ref{sct4}. Section~\ref{sct5}
summarizes the conclusions of the paper, while Sec.~\ref{sct6} is an addendum
with a response to comments by Stapp in the last section of \cite{ntk01}.

\section{Counterfactuals}
\label{sct2}

Suppose a gun fires a bullet which comes to a stop in a block of wood
separating the gun from a glass beaker.  What \emph{would have} happened
\emph{if} the wooden block had not been in the way?  This is a
\emph{counterfactual} question comparing an ``actual'' world $W_1$ in which
something took place, with a ``counterfactual'' world $W_2$ which resembles
$W_1$ in certain respects, but differs from it in others.  Such counterfactual
reasoning is quite common in everyday experience, but trying to codify it and
nail down the rules has given philosophers a lot of trouble.
A fundamental difficulty is that there are often several choices for $W_2$,
each different from $W_1$ but in different respects. If counterfactual
reasoning about the classical world of everyday experience involves
subtleties, one should not be surprised if similar or even worse difficulties
arise when trying to extend it to the quantum world.

Before discussing quantum issues let us complicate our classical example
slightly by assuming that the gun can be aimed in several different
directions, only one of which is towards the beaker, and the aim is determined
just before the gun is fired by tossing a coin.  This could be done by using
the random outcome of some quantum measurement, e.g., using the number of
radioactive atoms decaying during a specified time interval, though such a
quantum source of randomness is not essential for the present discussion.
Similarly, we might suppose that the wooden block is left in place or quickly
pushed away so that it no longer protects the beaker depending on the outcome
of a (suitably rapid) coin toss.  Our counterfactual question as to what would
have happened had the wooden block not been in the way now becomes less
definite, or at least could be interpreted in more than one way.  Should we be
thinking that the gun actually was directed at the beaker (and thus at the
wooden block) in the case of interest, or should we allow for the possibility
that it might have been pointed in some other direction?  The second
interpretation leads to a probabilistic answer. ``The beaker would have been
shattered with probability 1/4, and remained unbroken with probability 3/4''
is a sensible response if the gun could have pointed with equal probability in
four different directions, only one of them aimed towards the beaker.
Probabilistic answers to counterfactual questions can make perfectly good
sense in a stochastic setting.

In CQT Ch.~19 it is suggested that quantum situations analogous to the example
just discussed could be usefully analyzed in the following way.  First it is
necessary that all the properties, macroscopic or microscopic that one wishes
to discuss, whether in $W_1$ or $W_2$, must belong to a single quantum
framework or consistent family of histories.  This \emph{single framework
  rule} is fundamental to quantum reasoning according to the histories
approach, see Ch.~16 of CQT, so it is natural to require that it also be
observed when dealing with counterfactuals.  Sometimes this can be done by
thinking of a closed quantum system which contains a quantum coin toss (or
perhaps several tosses) with different outcomes corresponding to different
worlds.  Next, work backwards in time from what actually happens in $W_1$ to a
time at which $W_1$ is identical to $W_2$, and identify a state of affairs,
called the \emph{pivot}, which is the same in both worlds. Then apply the laws
of quantum dynamics to predict, generally in a probabilistic sense, what will
later take place in $W_2$.  It is often helpful to use a time-ordered tree
diagram as an aid to analyzing the situation; several examples, both classical
and quantum, will be found in Ch.~19 of CQT.

The result of this analysis will in general depend upon the framework of
consistent families used to define, and differentiate, $W_1$ and $W_2$, and on
the choice of the pivot.  Counterfactual questions can have a variety of
answers.  Another way of stating the matter is that such questions are often
to a certain degree ambiguous, and thus can be interpreted in different
ways. Choosing a specific framework and pivot makes the counterfactual
question more precise, but then the (probabilistic) answer to the question may
depend on this choice. In the previous example one might want to use a pivot
that corresponds to the situation just after the gun is fired, when the
direction in which it is aimed has already been decided, assuming $W_1$ and
$W_2$ are at this time identical.  In this case one can conclude that in the
counterfactual world, with the wooden block out of the way, the bullet will
certainly, with probability 1, shatter the beaker.  However, if the pivot is
chosen at a time before flipping the coin that determines the direction in
which the gun is aimed, the answer, as noted above, is only probabilistic,
and thus less definite.
To be sure, one might think it more natural, and perhaps closer to everyday
usage, if a ``sharper'' answer to a counterfactual question is preferable to a
less definite one.  In the situation at hand, ``the beaker would have been
shattered'' is sharper, more definite, than the less definite probabilistic
response.  In any case nothing is lost by focusing attention on the precise
respects in which the worlds of interest differ, and the route that leads to a
particular answer to a counterfactual question.

\section{Hardy's State}
\label{sct3}

In what follows it is helpful to employ an explicit wave function of the Hardy
\cite{Hrdy93} type.  In place of Hardy's original notation we use one that is
by now fairly familiar in discussions of quantum information.  Think of two
spin-half particles (qubits) $a$ and $b$, where for each particle the
orthonormal basis $\ket{0},\ket{1}$ consists of eigenkets of $S_z$,
\begin{equation}
  S_z\ket{0} = (1/2) \ket{0},\quad
 S_z\ket{1} = -(1/2) \ket{1},
\label{eqn1}
\end{equation}
in units of $\hbar$.
The alternative orthonormal basis defined by
\begin{equation}
 \st\, \ket{+} = \ket{0}+\ket{1},\quad
  \st\, \ket{-} = \ket{0}-\ket{1},
\label{eqn2}
\end{equation}
consists of eigenkets of $S_x$ with eigenvalues $+1/2$ and $-1/2$,
respectively.  In this notation we write the Hardy state in the form
\begin{equation}
  \sqrt{3}\ket{\psi_0} = 
\ket{0}_a\ot \ket{0}_b + \ket{0}_a\ot \ket{1}_b + \ket{1}_a\ot \ket{0}_b
\label{eqn3}
\end{equation}
(The reader may prefer to abbreviate $\ket{0}_a\ot \ket{1}_b$ to $\ket{01}$.)
The $a$ and $b$ subscripts are used both to label the states and to
differentiate the angular momentum operators for the two particles in an
obvious way: $S_{ax}$, $S_{bz}$, etc.  Once the Hardy state has been created,
by whatever means, at a time $t_0$ it remains unchanged under unitary time
development---a trivial time development operator $I_a\ot I_b$, corresponding
to zero magnetic field, so the spins do not precess---up to a time $t_2$ when
one particle or the other begins to interact with some measuring device.

In the histories approach measurement apparatus must, at least in principle,
be described in fully quantum mechanical terms. The proper way to do this in
the case of idealized measuring processes is discussed extensively in CQT, see
in particular Chs.~17 and 18, and more briefly in \cite{Grff11b}.  For
present purposes we may assume that an apparatus designed to measure $S_{bz}$
for particle $b$ is initially in the state $\ket{Z_b}$.  The unitary time
transformation describing the result of its interacting with particle $b$ in
the time interval from $t_2$ to $t_3$ is given by
\begin{equation}
  \ket{0}_b\ot \ket{Z_b} \ra \ket{0}_b\ot \ket{Z_b^+},\quad
 \ket{1}_b\ot \ket{Z_b} \ra \ket{1}_b\ot \ket{Z_b^-},
\label{eqn4}
\end{equation}
where one should think of $\ket{Z_b^+}$ and $\ket{Z_b^-}$ as macroscopically
distinct, corresponding to different pointer positions in the traditional
language of quantum foundations. The projectors on such states---equivalently,
the corresponding properties---will be denoted by $Z_b^+$ and $Z_b^-$.  One
can make the model more realistic in various ways, e.g., using density
operators for the apparatus, see Chs.~17 and 18 of CQT, but that is not needed
for the following discussion.  In a similar way an apparatus designed to
measure $S_{bx}$ for particle $b$ starts in a state $X_b$, and the time
transformation analogous to \eqref{eqn4} for the time interval from $t_2$ to
$t_3$ is
\begin{equation}
  \ket{+}_b\ot \ket{X_b} \ra \ket{+}_b\ot \ket{X_b^+},\quad
 \ket{-}_b\ot \ket{X_b} \ra \ket{-}_b\ot \ket{X_b^-}.
\label{eqn5}
\end{equation}

An apparatus designed to measure $S_{bz}$ can be changed into one that
measures $S_{bx}$ by turning on a suitable magnetic field just ahead of the
entrance of the $S_{bz}$ device, one which causes states $S_{bx}=\pm1/2$ to
precess into states $S_{bz}=\pm1/2$.  Let us refer to these two possibilities
as \emph{measurement settings} of the apparatus, denoted by $\ket{Z_b}$ and
$\ket{X_b}$ as used above.  We can imagine that the presence or absence of
this magnetic field is determined by a quantum coin of the sort described in
CQT, at a time just before the arrival of the $b$ particle.  In the same way
one can imagine that measurements of $S_{az}$ or $S_{ax}$ on particle $a$ can
be determined by an apparatus with settings $\ket{Z_a}$ and $\ket{X_a}$, which
could be set by a different quantum coin shortly before the arrival of the $a$
particle.

Table I relates the notation used here to that of Hardy \cite{Hrdy93} and
Stapp \cite{ntk01}. Thus $Z_a$ corresponds to Hardy's $U_1$ and Stapp's $\ML1$,
whereas the outcome $Z_a^+$ corresponds to $U_1=0$ and $\ML1+$, and $Z_a^-$ to
$U_1=1$ and $\ML1-$.

\begin{table}[h]
\caption{Comparison of notation}
$$
\begin{array}{ r@{\hspace{3pt}} l r@{\hspace{3pt}} l r@{\hspace{3pt}} l }
\multicolumn{2}{c}{\text{This paper}}&
\multicolumn{2}{c}{\text{Hardy}}&
\multicolumn{2}{c}{\text{Stapp}}\\
 Z_a\coln& Z_a^+, Z_a^-  & U_1\coln& U_1 = 0,1 & \ML1\coln& \ML1+, \ML1- \\
 X_a\coln& X_a^+, X_a^-  & D_1\coln& D_1 = 0,1 & \ML2\coln& \ML2-, \ML2+ \\
 Z_b\coln& Z_b^+, Z_b^-  & U_2\coln& U_2 = 0,1 & \MR1\coln&  \MR1-, \MR1+ \\
 X_b\coln& X_b^+, X_b^-  & D_2\coln& D_2 = 0,1 & \MR2\coln& \MR2+, \MR2- \\
\end{array}$$
\label{tbl1}
\end{table}

\section{Stapp's Counterfactual}
\label{sct4}

Stapp's counterfactual SR \cite{ntk01} translated into the notation of
Sec.~\ref{sct3} reads as follows:

\begin{quote}

  SR: Suppose that $S_{bz}$ is measured on particle $b$ and the result is
  $Z_b^-$.  Then if instead of $S_{bz}$, $S_{bx}$ had been measured on
  particle $b$ the result would have been $X_b^+$ with certainty, i.e.,
  probability one.
\end{quote}

One can derive SR using the approach to quantum counterfactuals in Ch.~19 of
CQT \emph{if} an appropriate choice is made for both framework and pivot.  A
suitable framework is the collection of histories
\begin{equation}
  [\Psi_0]\od \{[0]_a,[1]_a\}\od \{X_b,Z_b\}\od \{X_b^+,X_b^-,Z_b^+,Z_b^-\},
\label{eqn6}
\end{equation}
with time increasing from left to right, interpreted as follows. As in CQT we
use the notation $[\psi] = \dya{\psi}$ for the projector onto a quantum state
$\ket{\psi}$.  At the initial time $t_0$ the quantum state is
$\ket{\Psi_0}=\ket{\psi_0}\ot\ket{M}$, where $\ket{\psi_0}$ is the Hardy state
\eqref{eqn3} and $\ket{M}$ is the initial state of the apparatus which will
later measure some component of the spin of particle $b$, along with the
quantum coin which will determine whether the apparatus is set up to measure
$S_{bz}$ or $S_{bx}$. The $\od$ is a tensor
product symbol, but for present purposes one can regard it as simply
separating events at different times.  At $t_1$ particle $a$ is in one of the
two mutually exclusive states $\ket{0}_a$ or $\ket{1}_a$, denoted by the
projectors $[0]_a$ and $[1]_a$, and nothing is said about particle $b$ or the
apparatus.  As usual, $[0]_a$ is equivalent to $[0]_a\ot I_b\ot I_M$, where
$I_b$ and $I_M$ are the identity operators on particle $b$ and the
apparatus. At a later time $t_2$ the quantum coin flip results in the
apparatus being in one of the two states $X_b$ or $Z_b$, ready to measure
$S_{bz}$ or $S_{bx}$, whereas at a still later time $t_3$ the measurement
outcomes (pointer positions) are $X_b^+$, etc., in an obvious notation.

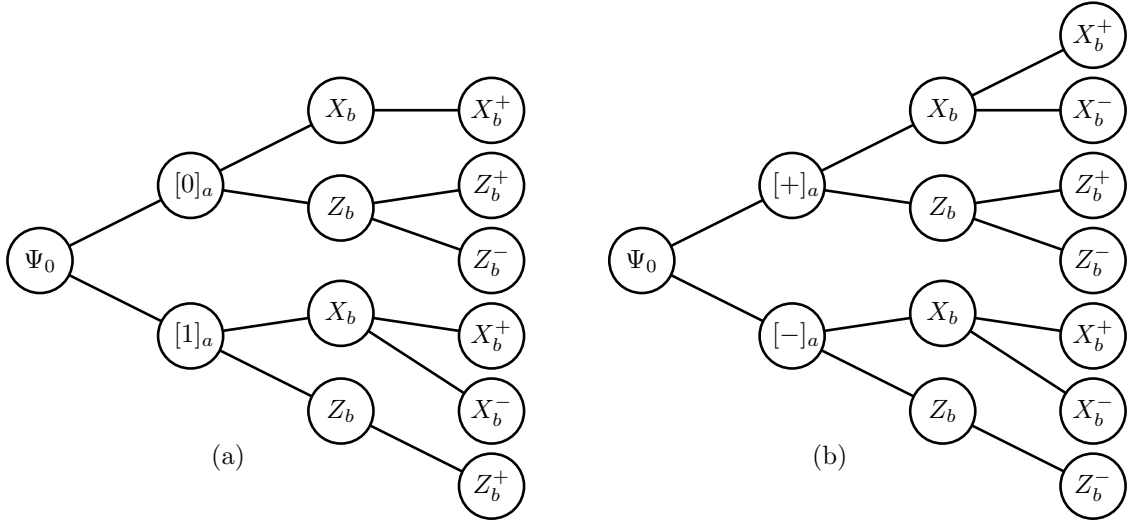
\begin{figure}[h]
$$
\begin{pspicture}(-4.5,-3.6)(10.5,3.6) 
\newpsobject{showgrid}{psgrid}{subgriddiv=1,griddots=10,gridlabels=6pt}
\def\lwd{0.035} 
\def\lwb{0.10}  
\def\lwn{0.01}  
\psset{
labelsep=2.0,
arrowsize=0.150 1,linewidth=\lwd}
\def\circb{
\pscircle[fillcolor=white,fillstyle=solid]{0.45}}
\def\dput(#1)#2#3{\rput(#1){#2}\rput(#1){#3}}
\def\nzz{0,0}
\def\naa{2,1}\def\nab{2,-1}
\def\nba{4,2}\def\nbb{4,0.7}
\def\nbc{4,-0.7}\def\nbd{4,-2}
\def\ncz{6,3}
\def\nca{6,2}\def\ncb{6,1}\def\ncc{6,0}
\def\ncd{6,-1}\def\nce{6,-2}\def\ncf{6,-3}
              \def\fga{
\psline(\nzz)(\naa) \psline(\nzz)(\nab)
\psline(\naa)(\nba) \psline(\naa)(\nbb)
\psline(\nab)(\nbc) \psline(\nab)(\nbd)
\psline(\nba)(\nca) 
\psline(\nbb)(\ncb)\psline(\nbb)(\ncc)
\psline(\nbc)(\ncd)\psline(\nbc)(\nce)
\psline(\nbd)(\ncf)
\dput(\nzz){\circb}{$\Psi_0$}
\dput(\naa){\circb}{$\;[0]_a$}
\dput(\nab){\circb}{$\;[1]_a$}
\dput(\nba){\circb}{$X_b$}
\dput(\nbb){\circb}{$Z_b$}
\dput(\nbc){\circb}{$X_b$}
\dput(\nbd){\circb}{$Z_b$}

\dput(\nca){\circb}{$X_b^+$}
\dput(\ncb){\circb}{$Z_b^+$}
\dput(\ncc){\circb}{$Z_b^-$}
\dput(\ncd){\circb}{$X_b^+$}
\dput(\nce){\circb}{$X_b^-$}
\dput(\ncf){\circb}{$Z_b^+$}
\rput[B](2.5,-2.7){(a)}
             }
              \def\fgb{
\psline(\nzz)(\naa) \psline(\nzz)(\nab)
\psline(\naa)(\nba) \psline(\naa)(\nbb)
\psline(\nab)(\nbc) \psline(\nab)(\nbd)
\psline(\nba)(\nca) \psline(\nba)(\ncz)
\psline(\nbb)(\ncb) \psline(\nbb)(\ncc)
\psline(\nbc)(\ncd) \psline(\nbc)(\nce)
\psline(\nbd)(\ncf)
\dput(\nzz){\circb}{$\Psi_0$}
\dput(\naa){\circb}{$\;[+]_a$}
\dput(\nab){\circb}{$\;[-]_a$}
\dput(\nba){\circb}{$X_b$}
\dput(\nbb){\circb}{$Z_b$}
\dput(\nbc){\circb}{$X_b$}
\dput(\nbd){\circb}{$Z_b$}

\dput(\ncz){\circb}{$X_b^+$}
\dput(\nca){\circb}{$X_b^-$}
\dput(\ncb){\circb}{$Z_b^+$}
\dput(\ncc){\circb}{$Z_b^-$}
\dput(\ncd){\circb}{$X_b^+$}
\dput(\nce){\circb}{$X_b^-$}
\dput(\ncf){\circb}{$Z_b^-$}
\rput[B](2.5,-2.7){(b)}
             }
\rput(-4,0){\fga}
\rput(4,0){\fgb}
\end{pspicture}
$$
\caption{%
Diagrams for (a) the family \eqref{eqn6}; (b)  the family \eqref{eqn7}.
}
\label{fgr1}
\end{figure}

The family of histories in \eqref{eqn6}, which is easily shown to be
consistent using the methods in CQT, can also be conveniently represented in
Fig.~\ref{fgr1}(a), where again time increases from left to right and the
different lines correspond to the different histories which occur with finite
probability.  Thus the line from $\Psi_0$ that terminates in the second node
from the top, labeled $Z_b^+$, in the right column indicates a history in
which, starting from the initial state, the $a$ particle was at time $t_1$ in
$\ket{0}_a$, $S_{az}=+1/2$, the quantum coin flip resulted in $Z_b$, thus
$S_{bz}$ and not $S_{bx}$ being measured, with an eventual measurement outcome
of $Z_b^+$. The other lines have analogous interpretations.  Two of the
nodes one might have expected in the right hand column are absent:
following $[0]_a$ and $X_b$ there is no $X_b^-$, and following $[1]_a$ and
$Z_b$ there is no $Z_b^-$. This is because the probabilities of these
histories are zero: they never occur.

Using Fig.~\ref{fgr1}(a) based on \eqref{eqn6} it is possible to derive the
counterfactual statement SR in the following way.  Suppose $S_{bz}$ is
measured and the outcome is $Z_b^-$.  The $Z_b^-$ node only occurs in the
upper half of the diagram, and tracing this history backwards in time one
concludes that at an earlier time particle $a$ was in the state $[0]_a$.
Using this node as the pivot, and assuming that in the counterfactual world
the quantum coin toss resulted in $X_b$ rather than $Z_b$, follow the line
from $[0]_a$ to $X_b$ and then on to $X_b^+$. The conclusion is that had
$S_{bx}$ been measured the result would have been $X_b^+$ with certainty,
probability one.

However, \eqref{eqn6} is not the only framework or family of histories one
might use to analyze the situation. An alternative family
\begin{equation}
  [\Psi_0]\od \{[+]_a,[-]_a\}\od \{X_b,Z_b\}\od \{X_b^+,X_b^-,Z_b^+,Z_b^-\},
\label{eqn7}
\end{equation}
differs from \eqref{eqn6} only in that the possible properties of particle $a$
at time $t_1$ are now the eigenstates of $S_{ax}$ instead of $S_{az}$.  As the
projectors associated with $S_{ax}$ do not commute with those associated with
$S_{az}$ the two families in \eqref{eqn6} and \eqref{eqn7} are incompatible:
the results of reasoning using one cannot be combined with results using the
other; this is the single framework rule discussed at length in CQT.  The
family \eqref{eqn7} corresponds to the diagram in Fig.~\ref{fgr1}(b), which
resembles (a), but differs from it in two respects, in addition to the fact
that the $[0]_a$ and $[1]_a$ nodes have been replaced with $[+]_a$ and
$[-]_a$.  First, the node $Z_b^-$, the starting point of the counterfactual
SR, occurs \emph{twice} at the final time $t_3$; it follows both $[+]_a$ and
$[-]_a$.  Therefore one cannot conclude from $Z_b^-$ that $S_{ax}$ had a
particular value at $t_1$: it could have been either positive or negative.
Second, both of the $X_b$ nodes at time $t_2$ are followed by both possible
outcomes, $X_b^+$ and $X_b^-$, with finite probabilities.  Either of these
changes is enough to prevent the derivation of SR in its precise form, given
above, if one uses the family \eqref{eqn7}.  One would, instead, have to be
content with a weaker form in which SR would end with: ``\dots the result
would have been $X_b^+$ with a probability $p$.''  Here $p$ is some
probability less than 1.

Thus the strict counterfactual SR (conclusion with probability 1) can be
derived using a certain consistent family, but only a weaker version can be
obtained using a different family.  Which is correct?  In the quantum world,
no less than in the classical world, there is in general no ``right'' way to
discuss counterfactuals.  Would the beaker have been shattered if the board
had not been in place?  The answer depends, as discussed in Sec.~\ref{sct2},
on how one goes about addressing the question, which is to say how one makes
it more precise by specifying the ways in which world $W_2$ coincides with and
how it differs from $W_1$.  That there is a particular framework in which SR
(in its original strict form) can be derived is a nontrivial results that
depends, among other things, on the form of the Hardy state \eqref{eqn3}. One
could adopt the convention that SR is correct provided one can find \emph{at
  least one} framework and pivot which justifies it, and in that case it
follows from the argument based on \eqref{eqn6} and Fig.~\ref{fgr1}(a); the
existence of other frameworks or pivots that lead to less definite results is
irrelevant.  The classical example analyzed in Sec.~\ref{sct2} would suggest
that such a convention is not implausible, but it would in any case be a
convention; there are no strict and universally accepted rules for how to
handle counterfactual arguments.  Stapp, in particular, does not accept this
convention \cite{ntk03}.

The preceding discussion employs \emph{properties} of particle $a$ and makes
no reference to \emph{measurements} on particle $a$. In contrast to textbook
quantum mechanics, the histories approach allows statements to be made about
microscopic properties without reference to measurements.  However, the
arguments we have just presented, and the diagrams in Fig.~\ref{fgr1}, remain
perfectly valid if one allows particle $a$ to interact at some time later than
$t_1$ with an apparatus (which must, of course, be included in the analysis as
part of the total quantum system) whose setting of $\ket{Z_a}$ or $\ket{X_a}$,
to measure $S_{ax}$ or $S_{az}$, may be chosen by a quantum coin near this
apparatus just before the arrival of particle $a$.  As neither of the families
in \eqref{eqn6} or \eqref{eqn7} makes any reference to the later measurement
on particle $a$ or its outcome, the probabilities employed in constructing
them, and therefore the diagrams in Fig.~\ref{fgr1}(a) and (b) remain
unchanged.  So again there is at least one framework in which SR can be
derived, and another (in fact many others) in which only a weakened form, with
probability $p<1$, can be obtained.  If one adopts the convention (which, as
noted above, Stapp rejects) that the sharpest answer to a counterfactual
question, the most precise one that can be found considering different
frameworks and pivots, is the correct one, SR is always correct, independent
of the outcome of the quantum coin flip that results in $\ket{X_a}$ or
$\ket{Z_a}$.  In particular the families in \eqref{eqn6} and \eqref{eqn7}
remain consistent if the set of events at time $t_2$ is augmented so that
$\{X_b,Z_b\}$ is replaced with the four possibilities $\{X_aX_b,
X_aZ_b,Z_aX_b, Z_aZ_b\}$.  (We remind the reader that the product $PQ$ of two
commuting projectors $P$ and $Q$ is to be interpreted as the conjunction of
the properties: $P$ \AND\ $Q$.)

But what can one say about measurement \emph{outcomes} on the $a$ side?  Here
care is needed: it is not possible to add the outcomes $X_a^-,X_a^+$ for
particle $a$ to the family \eqref{eqn6}, nor is it possible to add the
outcomes $Z_a^-,Z_a^+$ to the family in \eqref{eqn7}, without violating
consistency conditions and thus rendering the family meaningless.  One can, on
the other hand, achieve consistency in both cases either by saying nothing
(use the quantum identity operator) or by using appropriately chosen
macroscopic quantum superpositions---MQS or Schr\"odinger cat state---in place
of the pointer basis.  The reader is referred to the discussion in Sec.~19.4
of CQT, e.g., that associated with Eq.~(19.12), and we leave it as an exercise
to work out which MQS and which ``ordinary'' or ``pointer'' states are
permitted by the consistency conditions in the situation under consideration
here.  It should be emphasized that the histories approach does \emph{not}
have an automatic commitment to using the pointer basis: while in many
situations it is the most useful description to employ in terms of what one is
interested in discussing, this is a matter of utility, not necessity.

Thus whether or not one can derive SR (in the precise, not the weakened, form)
is very much a matter of the choice of framework and pivot.  There is always
at least one framework and pivot for which the such a derivation is possible,
and this is true for either outcome of the quantum coin flip that determines
the measurement settings $\ket{Z_a}$ or $\ket{X_a}$ for the apparatus that
will interact with particle $a$.  Stapp has made a particular choice of
framework, one might call it a ``hybrid,'' in which the derivation of SR is
possible when the coin results in $\ket{Z_a}$ but not when it results in
$\ket{X_a}$.  However, there is an alternative choice of framework, in which
the pointer basis is replaced by suitable MQS states, for which in the
$\ket{Z_a}$ case the derivation of SR is impossible, but is possible in the
case $\ket{X_a}$!  From the perspective of fundamental quantum theory there is
no reason to prefer one of these frameworks to the other.

Some additional points. First, the dependence of counterfactual conclusions on
the framework and pivot was pointed out explicitly in CQT, both in Ch.~19
where counterfactuals are first discussed, and in Ch.~25 where they are
applied to the paradox presented in an earlier paper by Hardy \cite{Hrdy92}.
It is not an idea developed more recently in order to respond to \cite{ntk01}.
Second, the choice of framework (and pivot) is one made by the physicist in
constructing a description of the quantum world, and is not some sort of
``physical influence'' upon that world.  See the discussion in Sec.~27.3 of
CQT.  A quantum framework is something like a point of view.  A coffee cup
looks different when viewed from below than when viewed from above.  Changing
the viewpoint does not change the cup, though it changes what one knows or can
say about the cup.
Third, as long as there is no interaction between particles $a$ and $b$
following their preparation, the temporal ordering of events on the $a$ (left)
side relative to those on the $b$ (right) side---e.g., whether measurements
are carried out first on particle $a$ or first on particle $b$---makes no
difference whatsoever for the consistency of the different families or the
probabilities of events within one family.  One can even assume that the
measurements on $b$ occur inside the backward light cone of the measurements
on $a$, or vice versa; see the discussion of such situations in
\cite{Grff11b}.  This is precisely what one would expect to be the case in the
\emph{absence} of any nonlocal influences of the sort that Stapp claims exist,
so it is an additional confirmation of their nonexistence. 

\section{Conclusion}
\label{sct5}

What we have shown is that, given the Hardy state and appropriate settings for
measurements on particle $b$, there are frameworks, with \eqref{eqn6} an
example, in which it is possible to use the method of counterfactual reasoning
presented in Ch.~19 of CQT in order to derive Stapp's SR in its strict form
(probability 1), while there are other frameworks, for example \eqref{eqn7},
in which such a derivation is not possible, and one can only obtain a weaker,
probabilistic form of SR.  The example given by Stapp in \cite{ntk01} is a
combination of the two, a framework in which SR can be derived provided a
quantum coin results in a measurement setting $\ket{Z_a}$ for the distant
particle $a$ and one uses a pointer basis for the measurement outcomes,
whereas the derivation is not possible if the coin leads to the setting
$\ket{X_a}$ and one again employs the pointer basis.
However, this fails to prove the existence of nonlocal influences in quantum
mechanics because, as discussed in Sec.~\ref{sct4}, there are alternative
frameworks in which one can arrive at different results. In particular,
there is a framework in which the case $\ket{Z_a}$ followed by an MQS basis
makes the derivation of SR impossible, whereas in the case $\ket{X_a}$ one can
derive SR.  The existence of these different frameworks undermines any
argument for nonlocality which depends upon a particular choice among them.  


Thus what Stapp has shown is not the existence of quantum nonlocality, but
instead the inadequacy of counterfactual reasoning of the sort he is
advocating for analyzing situations of this kind.  Fortunately, there are
other approaches to answering the question of nonlocal influences.  Neither
Stapp nor anyone else has yet found a defect in the relatively straightforward
(no counterfactuals) demonstration of the principle of Einstein locality given
in \cite{Grff11}, a principle which directly contradicts Stapp's claims of
nonlocality. 

None of this should be taken to imply that the study of counterfactual
reasoning in the quantum domain is impossible or uninteresting or must always
lead to ambiguous results.  Instead, it is a tool that needs to be used
carefully, with full recognition of its ambiguities and the possibility that
it can mislead, especially if one employs inconsistent ideas, such as
the treatment of measurement using wave function collapse found in current
textbooks, rather than starting off with a sound formulation of quantum
mechanics that is consistent with the mathematical structure of Hilbert space.

\section{Final Note}
\label{sct6}

The preceding sections were written before seeing the final section,
``Response to Griffiths' Reply'', in \cite{ntk01}. Hopefully the following
remarks will enable the reader to better understand the points at which Stapp
and I differ, and assess the merits of each position. Before mentioning the
differences it is worth emphasizing that there are significant points on which
we agree. Stapp prefers to use a framework which only includes macroscopic
outcomes of measurements, and he is free to do so. His conclusions drawn from
this frameworks about the correctness, or let us say the derivability, of the
counterfactual SR, are in accord with the inference scheme given in Ch.~19 of
CQT.  The same is true of the frameworks I prefer (and am free) to use, those
shown in Fig.~\ref{fgr1}, which include microscopic properties of particle $a$
but make no reference to measurements on this particle or their outcomes.  (It
is typical of the histories approach that different frameworks can be
employed, and there is no law of nature which specifies the ``correct''
framework. Frameworks are chosen on the basis of what issues one wants to
discuss, and there is a very general argument, Ch.~16 of CQT, for the overall
consistency of the histories approach provided one pays strict attention to
its single framework rule, which forbids combining incompatible frameworks.)

Our major disagreement is over the conclusions which can be drawn from these
analyses.  Stapp believes that because he has identified a framework which
properly corresponds to his earlier argument for nonlocal influences, and in
this framework the ability to deduce SR is linked to which measurement is
carried out on particle $a$, this demonstrates a nonlocal influence on
particle $b$. I disagree, because there exist alternative frameworks in which
there is no such link between measurement choices on $a$ and the derivation of
SR for $b$. The existence of alternative frameworks in which one can draw
different conclusions is already present in the case of classical
counterfactuals, as discussed in Sec.~\ref{sct2}, and one cannot expect the
quantum situation to be any simpler.  So why believe a conclusion found in one
framework but not supported by a similar analysis in another?  The reader will
have to judge whether Stapp makes a convincing case for nonlocality
or whether, as is my opinion, he has only demonstrated the hazards involved in
trying to use this mode of counterfactual reasoning to reach sound conclusions
about the quantum world.

Another point of agreement is that my proof of Einstein locality in
\cite{Grff11} is correct about what it asserts when it is applied to the
present situation in the following way: Let there be a third particle $c$
which is initially in one of two states $\ket{c}=\ket{0}$ or $\ket{1}$.  Let
it interact with the measuring apparatus associated with particle $a$
\emph{after} that particle has been separated from particle $b$, in particular
after the preparation of both particles in the entangled Hardy state
\eqref{eqn3}, in such a way that if $\ket{c}=\ket{0}$ the apparatus will
measure $S_{az}$, and if $\ket{c}=\ket{1}$ the apparatus will measure
$S_{ax}$.  Then the probabilities of any sequence of events involving the
distant particle $b$ or its measurement apparatus will be \emph{exactly the
  same} whatever the initial state, $\ket{0}$ or $\ket{1}$, of $c$.

Our disagreement is about whether Einstein locality in this form rules out any
nonlocal influence by the choice of measurement (as determined by $c$) on
particle $a$, on the distant particle $b$.  My position is that since this
measurement choice has no effect upon the probabilities which describe $b$ or
its associated measurement apparatus, it has no ``influence'' in any sense
akin to the usual notion of physical influence.  Stapp asserts that the
influence is not one that is revealed by probabilities, but by something else.
I confess I do not understand his reasoning at this point, so must leave it to
the reader to assess its validity.  He also asserts that the omission of a
counterfactual discussion from my derivation of Einstein locality represents a
severe deficiency, a fatal flaw.  Here again we disagree; Stapp has much more
confidence in the soundness of his counterfactual approach than do I. So again
the reader must judge.  Let me add that while the basics of the histories
approach as presented in CQT seem fundamentally sound---or at least no
significant flaws have thus far been pointed out by critics \cite{ntk06}---the
part in Ch.~19 having to do with counterfactual reasoning is more tentative
than the rest. In particular, it involves the direction (past versus future)
of time in a way not present in the main formalism, which is time symmetric.
Revisions of this aspect of the histories approach to quantum counterfactuals,
which has a very definite time asymmetry, may be needed when the problem of
thermodynamic irreversibility has been better understood in quantum terms.

\section*{Acknowledgments}

I thank Henry Stapp for providing a preliminary version of \cite{ntk01}, for
several rounds of correspondence which forced me to clarify my thinking about
quantum counterfactuals, and for comments on a draft version of this
manuscript. The research described here received support from the National
Science Foundation through Grants PHY-0757251 and PHY-1068331.


\begin{thebibliography}{10}
\bibitem{ntk02}
For other work that comes to the same conclusion, see
  \cite{Grff87,Grff02b,Grff11b}.
\end{thebibliography}
\end{document}